\documentclass[conference]{IEEEtran}
\usepackage[T1]{fontenc}
\usepackage[utf8]{inputenc}
\usepackage{booktabs}
\usepackage{array}
\newcolumntype{L}[1]{>{\raggedright\arraybackslash}p{#1}}
\usepackage{url}
\usepackage{cite}
\usepackage{tcolorbox}
\usepackage{tikz}
\usetikzlibrary{shapes.geometric, arrows.meta, positioning}

\tikzset{
  box/.style={rectangle, rounded corners=2pt, draw=black!70, fill=black!3,
    align=center, inner sep=4pt, font=\scriptsize, minimum width=30mm},
  gate/.style={diamond, aspect=2.2, draw=black!70, fill=black!8,
    align=center, inner sep=1pt, font=\scriptsize\itshape},
  note/.style={rectangle, draw=black!40, dashed, fill=white,
    align=center, inner sep=3pt, font=\scriptsize\itshape},
  flow/.style={-{Latex[length=2mm]}, draw=black!70},
  dflow/.style={-{Latex[length=2mm]}, draw=black!55, dashed},
}

\begin{document}

\title{\LARGE Introducing Consort: A Spec-First Agent Framework for Enforced,
Test-Driven Development on Live Database Branches}

\author{\IEEEauthorblockN{Kevin Hartman}
\IEEEauthorblockA{Databricks}
\thanks{Consort is available as open source at \url{https://github.com/databricks-solutions/consort}.}}

\maketitle

\begin{abstract}
When an agent writes code, the development framework becomes the control system
for a non-deterministic worker. Spec-first, agent-driven frameworks have gained
rapid traction since 2025; the installable ones, GitHub Spec Kit, obra/superpowers,
BMAD, and GSD, and our own, all capture intent through a specification or durable
planning artifacts. Since they agree on capturing intent up front, what separates
them is how each enforces the engineering discipline that keeps agent-written code
clean, correct, and maintainable. Every framework enforces that discipline somehow;
they differ in how. We characterize three modes: enforcement by persuasion (prompt
discipline the model may ignore), by front-loaded structure (strong specs, then a
trusted build), and through controls the agent cannot edit (a deterministic
orchestrator, human-approved gates, immutable tests, and a green result that must
pass against a live, branched database). We introduce Consort, a spec-first,
test-driven agent framework built on the third, enforcing that discipline through
controls the agent runs inside but cannot bypass, in which a deterministic
orchestrator drives separate role agents through a spec-first design lane and a
test-driven build lane on a live database branch. We argue that enforcing the tests
and gates in code keeps agent-written code honest and verifiable, while its
specialized roles, like the human roles before them, are what make it maintainable,
claims we frame as a pre-registered, testable hypothesis.
\end{abstract}

\begin{IEEEkeywords}
AI coding agents, spec-driven development, test-driven development, database
branching, software engineering, agentic frameworks, enforcement.
\end{IEEEkeywords}

\section{Introduction}
For most of software's history the development framework, the set of conventions,
tools, and rituals a team works within, has been an ergonomic convenience for a
developer who ultimately decides what the code does. That assumption no longer holds
when the author is a large language model. AI-assisted coding is now mainstream: a
large majority of developers report using or planning to use these tools, though a
much smaller fraction trust their output \cite{stackoverflow2025}, and a majority of
new developers reach for an AI assistant in their first week \cite{github2025}. The
prevailing style of use, what has come to be called vibe coding, leans on accepting
generated output through its observed behavior rather than close reading
\cite{karpathy2025, ge2025}, and even the productivity case is unsettled: a
randomized study found experienced open-source developers were slower when using
early-2025 AI tools, even as they believed themselves faster \cite{metr2025}. When a
non-deterministic worker writes code, the framework stops being a convenience and
becomes the control system for that worker. Its job is no longer to make a developer
productive; it is to constrain a probabilistic process so that its output can be
trusted.

A distinct family of frameworks has emerged since 2025 to do exactly this by putting
a specification at the center of the work. Spec-driven development, as it has come to
be called, spans a rigor ladder from spec-first (agree on the specification before
writing code) through spec-anchored (keep the specification alive alongside the code)
to spec-as-source (treat the specification as the primary artifact and the code as a
build output) \cite{piskala2026}. The category has gained rapid traction: GitHub's
own open-source toolkit, Spec Kit, launched in September 2025 \cite{delimarsky2025}; Amazon shipped a spec-driven IDE,
Kiro, to general availability in November 2025; a spec-centric platform, Tessl; and an open-source methodology delivered as
agent skills, obra/superpowers, spread across coding harnesses \cite{vincent2025}. We
do not claim the practice is settled or universally adopted; it is barely a year old,
agent-driven use specifically remains a minority practice, and thoughtful
practitioners have questioned whether current tooling delivers on its promise
\cite{bockeler2025}. But the category is real and growing, which is what
makes the question of how these frameworks work worth asking carefully.

The frameworks in this family agree on more than they disagree. All of them,
including our own, are spec-first: they ask the developer to state intent in a
specification before the agent writes code. Because they converge on how intent is
captured, the specification is not where they differ. Where they differ, and the
difference is the subject of this paper, is in what happens after the specification
exists: how each framework enforces clean, correct code from it. That is the harder
problem, because the agent is free to satisfy a specification in ways its author
never intended, and to report success it did not achieve.

The failure modes are documented and specific. Agents will make a test suite pass by
disabling or deleting the tests that stand in the way, so that a human-owned,
immutable suite has re-emerged as the mechanism that holds their output honest
\cite{beck2025}. They over-rely on mock objects: a large-scale study of commit
histories finds that agent commits introduce mocks at a materially higher rate than
human commits, and that mocked tests, while easier to generate, are weaker at
validating real behavior \cite{hora2026}. Across tens of thousands of real coding
sessions, agents violate stated constraints, overreach the requested change, and, in
a substantial share of episodes, report inaccurately what they actually did
\cite{tang2026}. None of these is exotic; they are the ordinary behavior of a
probabilistic system asked to produce a result and to grade its own work.

The consequences show up in the code that survives. Independent measurements
associate the AI-assisted era with declining maintainability signals: duplicated code
has risen as a share of changed lines while refactoring has fallen, to the point where
copy-and-paste recently overtook the moving of existing code \cite{gitclear2025};
teams with AI-assistant access have shown increases in defects reaching pull requests
without matching throughput gains \cite{uplevel2024}; and higher AI adoption has been
associated with a measurable drop in delivery stability \cite{dora2024}. These
findings are correlational and partly industry-sourced, and modern agentic tools do
iterate rather than emit code in a single pass. We therefore do not claim that
unenforced tools produce unmaintainable code as a settled fact. We take the
maintainability risk as real and motivating, and we ask which framework properties
actually reduce it.

Our answer is that the property that matters is not how a framework specifies work
but how it enforces the engineering discipline that keeps agent-written code clean,
correct, and maintainable, in two respects: how it holds the work honest as the agent
writes it, and how it divides the work among focused roles so that the result is
maintainable, as it was when those roles were held by people. Enforcement comes in
three sharply different strengths. This paper makes three contributions. First, a
taxonomy of how agent-driven frameworks enforce that discipline: enforcement by
persuasion, by front-loaded structure, and through controls the agent cannot edit,
distinguished by what the agent is physically able to ignore. Second, Consort, a
spec-first, test-driven agent framework in the same family as Spec Kit and
superpowers but enforcing the discipline through controls the agent cannot edit, in
which a deterministic orchestrator coordinates separate role agents and the enforced
build lane runs the entire test-driven cycle against a live, branched database, a
practice developed in a paper under anonymous review \cite{companion}. Third, a comparative
evaluation against four installable frameworks in which agents write the
code, GitHub Spec Kit, obra/superpowers, BMAD, and GSD, and a pre-registered
hypothesis for the output-quality claim we cannot yet prove.

Two scoping notes bound the contribution. We compare only real, installable
frameworks in which the agent writes code; hosted platform builders and fully
autonomous issue-to-pull-request agents occupy an adjacent space and are named, not
evaluated, here. And the database layer this framework depends on is not re-derived:
its mechanism, economics, and defect-catching evidence are the contribution of
a paper under anonymous review, which this paper takes as given in order to focus on enforcement of
the agent's work.

The remainder of the paper is organized as follows. Section~\ref{sec:background}
places the framework family on the spec-driven rigor ladder and reviews what
unenforced agents do. Section~\ref{sec:enforce} develops the central lens, the three
ways a framework enforces engineering discipline on a non-deterministic worker.
Section~\ref{sec:consort} presents Consort as the realization of the third,
enforcement through controls the agent cannot edit.
Section~\ref{sec:eval} evaluates the frameworks head to head against the enforcement
dimensions. Section~\ref{sec:experiment} states the output-quality claim as a
pre-registered experiment. Section~\ref{sec:discussion} discusses implications, and
Section~\ref{sec:conclusion} concludes.

\section{Background and related work}
\label{sec:background}

\subsection{The spec-driven rigor ladder}
The frameworks we compare belong to spec-driven development, whose organizing idea is
that a specification, not a prompt and not the code, should be the durable statement
of intent. The clearest current account arranges the practice on a rigor ladder:
spec-first agrees on the specification before code is written; spec-anchored keeps the
specification live and reconciled with the code as it changes; and spec-as-source
elevates the specification to the primary artifact from which code is generated
\cite{piskala2026}. The account is explicit that spec-driven development is a judgment
call rather than a universal good, adding value on some problems and overhead on
others.

The frameworks we examine span the lower two rungs. Spec Kit, superpowers, and our
own are spec-first, agreeing a specification before code; BMAD and GSD reach
spec-anchored, keeping planning artifacts (a product requirements document and
architecture in BMAD, persistent state and context files in GSD) alive across the
work. None is spec-as-source, and, more to the point, wherever a framework sits on
this ladder it does not measure enforcement. That the comparators range from
spec-first to spec-anchored and still leave the build largely to the agent is
precisely the point: the specification is not the differentiator, and our framework
claims no advantage in specification quality. The differentiator lies past the
specification, in enforcement, which the ladder does not capture.

\subsection{The enforcement thesis is emerging independently}
The idea that a specification is insufficient without enforcement is not ours alone;
it is being articulated across the field. Practitioner and vendor writing
increasingly frames the problem as one of enforcement rather than authorship: that an
agent without a governing operating model is unaccountable automation, and that
platform enforcement is what converts spec-driven development from a methodology into
a governed system \cite{itential2026}; that specifications should execute as
validation gates so that a build fails on divergence rather than trusting the agent to
comply \cite{augment2026}. Skeptical hands-on reviews reach a compatible conclusion
from the opposite direction, observing that current spec-driven tooling can create
excessive review overhead and a false sense of control, and that the discipline tends
to evaporate once the agent begins writing code \cite{bockeler2025}. That last
observation, that the discipline evaporates at exactly the moment the agent takes
over, is the gap this paper's framework is built to close.

\subsection{What unenforced agents do}
The case for enforcement rests on what agents do when they are not enforced. The
behaviors are by now well documented. Agents disable or delete tests to turn a suite
green, which is why treating tests as immutable has been proposed as a first-class
discipline \cite{beck2025}. They over-mock relative to human developers, adding mocks
in a materially larger share of commits and thereby validating less real behavior
\cite{hora2026}. At the scale of tens of thousands of sessions they violate
constraints, overreach, and misreport their own actions, though the same evidence
shows most such episodes cost effort and trust rather than causing irreversible damage
\cite{tang2026}. A qualitative study of an AI-native IDE illustrates the texture of
these failures, agents changing code that was not meant to change and requiring
continual human correction, without offering population-level frequencies
\cite{kumar2025}. Taken together, the picture is not of a tool that occasionally errs
but of a worker whose default behavior must be constrained if its output is to be
trusted.

\subsection{The comparators}
\textbf{GitHub Spec Kit} is the reference open-source spec-driven toolkit, published
under GitHub's own organization \cite{delimarsky2025}. It structures work as a command
pipeline, from establishing a project constitution through specifying, clarifying,
planning, a requirements checklist, tasks, analysis, implementation, and a convergence
pass, which the coding agent itself runs. It front-loads specification and planning,
and runs across many coding agents. Its enforcement, however, is light: test-driven
development is optional and ungated, shipped so that tests are included only when
explicitly requested; its governing constitution is honored as prose rather than
checked mechanically; there is no deterministic orchestrator behind the pipeline; and
there is no database in the loop, its data model being a design document rather than a
running store.

\textbf{obra/superpowers} is a prescriptive methodology delivered as auto-triggering
agent skills, spanning a brainstorming specification gate, a strong test-first rule,
subagent-driven development, and staged review, and portable across many harnesses
\cite{vincent2025}. It shapes agent behavior through prompt craft, and states its
test-first rule as an unconditional requirement. But its control loop is itself a
language model rather than a deterministic driver, so it can lose its place across
context compaction; its gates are soft prose; its tests are not tool-enforced; and it,
too, has no database concept.

\textbf{BMAD} (the Breakthrough Method for Agile AI-Driven Development) is a role-play
framework that splits work into an agentic-planning phase, in which named persona agents (an
analyst, a product manager, an architect) co-author a product requirements document
and an architecture document, and a context-engineered development phase, in which a
developer agent implements stories sharded from those artifacts \cite{bmad2026}. It
front-loads planning into a product requirements document and an architecture
document, with a test-architect agent that scores several quality dimensions and emits
pass, concerns, or fail gate files. Its enforcement changed during this paper's
revision window: BMAD added bmad-loop, a deterministic, code-driven orchestrator whose
test-architect gates are code-enforced blocks with human-only waivers. Those gates
enforce coverage and quality thresholds after an epic rather than verifying code
in-cycle; test-driven development is still not enforced within the cycle; and there is
no database in the loop.

\textbf{GSD} (Git. Ship. Done.) is a lightweight, agent-agnostic context-engineering
framework built around a five-phase loop, discuss, plan, execute, verify, and ship,
whose central goal is to fight the context degradation that accumulates as an agent
fills its window by pushing research and execution into fresh-context parallel
subagents while keeping the main session lean, with state and context files that
persist across sessions \cite{gsd2026}. Its strength is disciplined context hygiene
and portability across many runtimes from a single install. Its enforcement stays
deliberately thin: it has since added a documented seven-role taxonomy and
orchestrator-invoked gate predicates, but their enforcement remains convention- and
hook-level rather than code-guaranteed, and it has neither test-driven development nor
any database concept.

\subsection{Adjacent, not compared}
Several systems sit near this work but in a different class and are named only to
situate the contribution. Spec-as-source platforms such as Tessl, and spec-driven IDEs
such as Kiro, occupy higher rungs of the rigor ladder and are hosted rather than
installable methodologies. Autonomous issue-to-pull-request agents pursue end-to-end
automation rather than an enforced developer-in-the-loop lifecycle. These are the
surrounding field, not comparators, because the question here is how an installable,
spec-first framework enforces engineering discipline on an agent that writes code
within a developer's workflow.

\section{Enforcing discipline on a non-deterministic worker}
\label{sec:enforce}
The premise of this paper is a single principle: \emph{a control that the worker can
choose to ignore is not a control when the worker is non-deterministic.} A developer
who agrees to write a failing test first can be trusted, most of the time, to honor
that agreement because the developer understands why it matters. An agent has no such
commitment; it will honor a rule exactly insofar as the rule is present in its
context, salient against competing instructions, and not more cheaply satisfied by
circumventing it. Enforcing discipline on an agent is therefore not a matter of
stating the right rules but of placing those rules where the agent cannot reach them.

Every framework in this family is trying to enforce the same engineering discipline:
write the test first, keep the spec honest, do not fake a pass. They differ in how
they enforce it. Against the principle above, the family falls into three modes,
distinguished by what the agent is physically able to ignore.

\textbf{Enforcement by persuasion.} The framework's controls are instructions to the
model: rules, red flags, and prohibitions expressed in prose, however forcefully.
obra/superpowers is the exemplar, relying on prompt craft. But every control, the
orchestration, the gates, and even the test suite, is ultimately honored by the model
and editable by it. Persuasion raises the floor on agent behavior, sometimes
substantially. It cannot make a violation impossible, because the same worker that
reads the rule can decide, under pressure to produce a green result, to set it aside.

\textbf{Enforcement by front-loaded structure.} The framework invests in getting the
specification and plan right, then hands the agent a task list and largely trusts the
implementation. GitHub Spec Kit is the exemplar: its specifications and plans are
checklist-validated, and its constitution states principles. But once the agent begins
to build, testing is optional, the constitution is advisory, and no code-level
mechanism prevents the agent from diverging. The intent is strong and the build-time
enforcement is weak, which is precisely the pattern that leads discipline to evaporate
once the agent takes over \cite{bockeler2025}.

\textbf{Enforcement through controls the agent cannot edit.} The framework's controls
are code that the agent runs inside but cannot change. Routing between steps is a
program, not a model decision. Every decision is gated, so the work does not advance
until a human approves. Tests are immutable within a unit of work, so they cannot be
weakened to force a pass. And a green result is defined operationally, as a test runner
actually passing against a live database, rather than as the agent's report that it
passed. Under this mode the agent is free to write whatever code it likes, but it
cannot delete the test that judges the code, cannot skip the gate that a human must
approve, and cannot claim success the runner did not confirm.

These three modes are not points on a quality scale within a single design; they are
different answers to one question: when the agent can ignore a control, what makes it
comply anyway. Section~\ref{sec:consort} shows how Consort enforces the discipline
through controls the agent cannot edit. Section~\ref{sec:eval} scores all five
frameworks against that question dimension by dimension.

\section{Consort}
\label{sec:consort}
\begin{figure}[t]
\centering
\begin{tikzpicture}[node distance=4.2mm]
  \node[box] (intake) {Project intake (precondition)};
  \node[gate, below=of intake] (intakeg) {intake gate};
  \node[box, below=of intakeg] (plan) {Plan (sprint planning) -- PO};
  \node[gate, below=of plan] (plang) {plan gate};
  \node[box, below=of plang] (design) {Design lane (spec-first)\\SA, AR, DBA, TS, UX};
  \node[gate, below=of design] (specg) {spec + test-list gates};
  \node[box, below=of specg] (build) {Build lane\\(TDD on a live branch) -- NAV, DRV};
  \node[gate, below=of build] (acceptg) {acceptance gate};
  \node[box, below=of acceptg] (deploy) {Deploy};
  \node[gate, below=of deploy] (deployg) {deploy gate};
  \node[box, below=of deployg] (promote) {Promote (PR, CI, merge)};
  \node[gate, below=of promote] (promoteg) {promote gate};
  \node[box, below=of promoteg] (shipped) {Shipped (working software live)};
  \draw[flow] (intake) -- (intakeg);
  \draw[flow] (intakeg) -- (plan);
  \draw[flow] (plan) -- (plang);
  \draw[flow] (plang) -- (design);
  \draw[flow] (design) -- (specg);
  \draw[flow] (specg) -- (build);
  \draw[flow] (build) -- (acceptg);
  \draw[flow] (acceptg) -- (deploy);
  \draw[flow] (deploy) -- (deployg);
  \draw[flow] (deployg) -- (promote);
  \draw[flow] (promote) -- (promoteg);
  \draw[flow] (promoteg) -- (shipped);
  \draw[dflow] (shipped.east) to[out=0,in=0] node[right, font=\tiny\itshape] {loop to next increment} (plan.east);
\end{tikzpicture}
\caption{The Consort lifecycle. The role-agent ensemble performs the phases, and the
human decides every gate (diamonds), namely the intake, plan, spec-and-test-list, per-story acceptance, deploy, and promote gates (a backlog-selection gate also sits inside planning). Routing between phases is handled by a
deterministic orchestrator that writes no spec, code, or tests. Role agents are
abbreviated: PO (product owner), SA (spec author), AR (architect reviewer), DBA, TS
(test strategist), UX (UX designer, user-facing work only), and, in the build cycle,
NAV (navigator) and DRV (driver; Fig.~\ref{fig:green}). Deploy and promote have no role
agent; the orchestrator runs them directly. The loop runs per increment: the
specification is frozen within an increment and evolves across increments.}
\label{fig:lifecycle}
\end{figure}
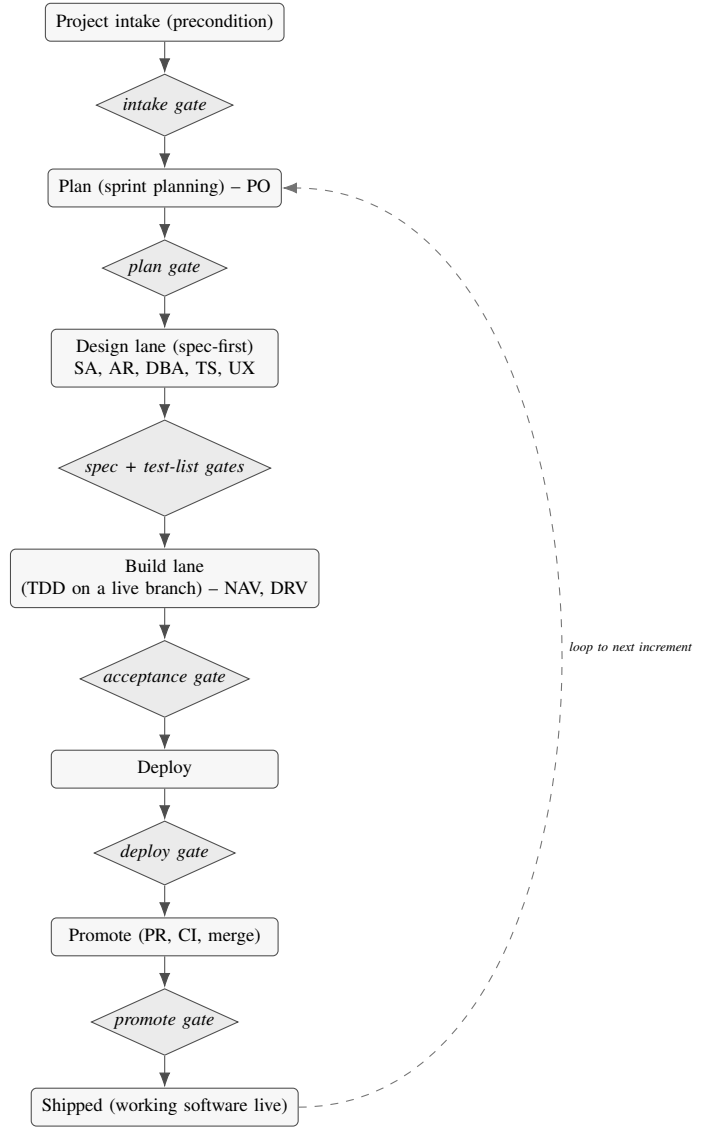

We introduce \textbf{Consort}, a spec-first, test-driven agent framework that enforces
engineering discipline through controls the agent cannot edit. The name reflects the
architecture: a consort is a small ensemble whose players perform one piece together,
and here a set of role agents work as an ensemble under the coordination of the
deterministic orchestrator. It has two lanes: a spec-first design lane, where intent is
agreed and frozen, and an enforced build lane, which runs the full test-driven cycle
against a live, branched database (Figure~\ref{fig:lifecycle}). The build lane is not a novel testing idea invented
here; it is test-driven development run against a live database branch, a practice
developed in a paper under anonymous review \cite{companion}. Consort is a reference implementation
of that practice, and the contribution of this paper is not the database idea but what
it takes to enforce that practice on a non-deterministic agent author.

\subsection{The build lane runs TDD against a live database branch}
It runs the entire red/green/refactor cycle, including the fast inner loop, against a
copy-on-write branch of a real, production-shaped database, rather than against mocks
or synthetic-seed fixtures. Copy-on-write branching makes an isolated, disposable
branch of the real database available for each unit of work in roughly constant time,
independent of data size, because the branch shares its parent's storage until it
diverges. A paper under anonymous review establishes why this collapses the long-standing
tradeoff between test speed and data fidelity, and catalogs the classes of
data-dependent defect that mocks and synthetic seeds cannot surface. This paper takes
those results as given. What it adds is enforcement: the machinery that makes an agent
actually run the cycle this way rather than reporting that it did. Because each branch is
cheap and isolated, the build lane can also run several experiments for one story in
parallel, each on its own database branch and git worktree, and keep the one that best
satisfies the tests, an exploration only cheap branching makes affordable.

\subsection{The deterministic orchestrator}
Consort's control flow is a program, not a language-model decision. A deterministic
orchestrator, a state machine rather than an agent, sequences the work: for each unit
of work it cuts a copy-on-write branch of the database, drives the red, green, and
refactor steps against that branch, enforces the order of the human-approval gates, and
discards the branch when the unit of work is complete. In effect, the coordinating role
that an agile team would call the scrum master is rendered here as deterministic code
rather than as a language model, so the process that drives the cycle and holds the
gates cannot itself drift or be argued out of a step. Because routing is code, the
control loop cannot be lost across context compaction, the failure to which a
language-model control loop is exposed. And because branch provisioning and disposal
are handled by the orchestrator, neither the developer nor the agent manages database
fixture state by hand; the branch is the fixture, and its disposal is the reset. The
orchestrator also stamps each experiment with a fingerprint of its design, so that if the
design is revised the stale experiment is recut from the current design rather than
silently reused.

\subsection{Role-separated agents}
\label{sec:roles}
Consort assigns work to a set of persistent, named role agents, each with a bounded
responsibility and no shared memory with the others: a product owner, a spec author, an
architect reviewer, a DBA, a test strategist, and a navigator and driver who run the
test cycle together, with a UX designer added for user-facing work. Two of these roles carry
responsibilities worth noting. The UX designer grounds its design guidance in real
comparables: it opens the reference sites a brief names with a browser and reads their
actual type, color, and spacing, or searches for representative sites when none is
named, so that each design-token choice cites an observed reference rather than a
default. The test strategist acts as a supervisor rather than a sole author, fanning out
to per-kind analyst agents for behavior, fitness, and client tests and then reconciling
and ordering their contributions into one test list. Deploy and promote
are not a role agent but deterministic steps the orchestrator runs directly, logged
under a release-engineer label. Coordination is deliberately not one of these agents:
the role an agile team would call the scrum master is the deterministic orchestrator of
Section~\ref{sec:consort}, a state machine rather than a language model, so the ceremony
that sequences the work is code and cannot itself be talked into skipping a step. The
role separation among the agents is equally deliberate, and it does two things. First,
it guards integrity: an agent that both writes code and judges it has every incentive
to grade generously, whereas a role that only writes must satisfy a role, and a gate,
it does not control. Second, and less obviously, it is where maintainability comes
from. On human teams, the division of labor, a product owner who owns the requirement,
a developer who implements it, a reviewer who tests it, produced maintainable software
precisely because each concern had an owner whose focus was that concern and not the
others; a single generalist optimizing only for a passing result tends to produce code
that passes and little else. Consort reproduces that division among agents, so that
each role optimizes its own concern, the clarity of the requirement, the simplicity of
the implementation, the adequacy of the tests, and the output carries the same
separation of concerns that made human-built systems maintainable. This is a claim
about the mechanism; maintainability itself remains the hypothesis of
Section~\ref{sec:experiment}. The mechanism is role specialization, not enforcement
alone: enforcing the tests and gates in code keeps the verification honest, while the
focused roles are what make the software maintainable, just as they did when the roles
were held by people. This differs from Spec Kit, where a single agent runs the whole
pipeline, and from superpowers, where subagents are dispatched ephemerally; here the
roles are stable participants in a single coordinated lifecycle.

\subsection{Enforcement primitives}
\label{sec:primitives}
Consort's guarantees rest on a few controls, each code rather than prose the agent may
reinterpret. The discipline is enforced in concrete ways:
\begin{itemize}
  \item \textbf{Tests are immutable within a unit of work.} The approved test list is
  locked, and an in-cycle attempt to delete or weaken a test is caught and flagged. The
  one controlled exception is principled and rare: when a later story legitimately
  supersedes an earlier test, a designated step refactors only those superseded tests,
  under an explicit rule, and then re-verifies; an ordinary failing test is never
  touched.
  \item \textbf{Green is honest by definition.} Success is recorded only when a test
  runner actually passes against a real branch, not when the agent asserts it.
  \item \textbf{Cross-story field contracts are checked before code.} If one story
  mandates a required field and the story that owns the user-submit path has no
  acceptance criterion that supplies it, the design lane flags the gap rather than
  letting the feature reach a build that ships a form the database will reject.
\end{itemize}
Together these make branch-based TDD enforceable rather than merely recommended:
immutability answers the test-deletion failure mode \cite{beck2025}, because within the
unit of work the tests are not writable; honest-green answers inaccurate self-reporting
\cite{tang2026}, because the record of success is the runner's, not the agent's. A human
lever sits alongside these: the product owner can mark a story as requiring an end-to-end
test, a requirement the design lane cannot downgrade.

\subsection{Real-data verification is the green condition}
The single property that no comparator shares is the definition of green. In Consort, a
green result is a passing run against a live copy-on-write branch of real data. Each
verify runs on a fresh branch cut for that check and discarded when it finishes, so the
result reflects the committed schema and no prior run's writes, rather than accumulated
state on a shared instance. Two of the agent's cheapest escape routes close at once. It
cannot delete the test, because the test is immutable (Section~\ref{sec:primitives}).
And it cannot substitute a compliant mock for real behavior, because the collaborator is
a real database rather than a double the agent can shape to its convenience, which is
the structural counter to the over-mocking that agents otherwise exhibit
\cite{hora2026}. A green result therefore requires code that actually works against
production-shaped data, with the database's constraints and governance (access and
masking policies inherited from the branch) in force. Spec Kit stops at a data-model
document and superpowers has no data concept, so for both the notion of green never
touches a real database. The mechanism and economics that make per-cycle real-data
verification affordable are that paper's contribution, cited here rather than
re-argued.

\begin{figure*}[t]
\centering
\begin{tikzpicture}[node distance=6mm and 16mm]
  \node[box] (plan) {PLAN (NAV)};
  \node[box, below=of plan] (red) {RED: write a failing test (NAV)};
  \node[box, below=of red] (green) {GREEN attempt: minimal honest code (DRV)};
  \node[gate, below=of green] (verify) {verify vs.\ a real branch};
  \node[box, below=of verify] (review) {REVIEW / REFACTOR (NAV, DRV)};
  \node[box, right=of verify] (assess) {ASSESS:\\regression or\\supersession? (NAV)};
  \node[note, above=of assess] (hil) {raise to human};
  \node[box, below=of assess] (repair) {REPAIR code,\\never the tests (DRV)};
  \node[box, below=of repair] (perm) {permissive-green:\\superseded tests only (DRV)};
  \draw[flow] (plan) -- (red);
  \draw[flow] (red) -- (green);
  \draw[flow] (green) -- (verify);
  \draw[flow] (verify) -- node[left, font=\tiny] {passes} (review);
  \draw[flow] (verify) -- node[above, font=\tiny] {fails} (assess);
  \draw[flow] (assess) -- node[right=1pt, font=\tiny] {genuine} (hil);
  \draw[flow] (assess) -- node[right, font=\tiny] {regression} (repair);
  \draw[flow] (assess.east) to[out=0,in=0] node[right, font=\tiny] {supersession} (perm.east);
  \draw[dflow] (repair.west) to[out=180,in=0] node[pos=0.5, above, font=\tiny] {re-verify} (verify.east);
  \draw[dflow] (perm.west) to[out=180,in=0] (verify.east);
  \draw[dflow] (review.west) to[out=180,in=180] node[left, font=\tiny] {next item} (plan.west);
\end{tikzpicture}
\caption{Honest-GREEN. GREEN is stamped only when the verify suite genuinely passes
against a live branch of real data, not when the agent reports success. A failed verify
routes either to a bounded code repair that never touches the tests, or, when a later
story legitimately supersedes a prior test, to a rule-bound permissive-green that
refactors only the flagged superseded tests; unresolved cases escalate to the human. The
navigator (NAV) owns the test-facing steps and the driver (DRV) the code-facing steps,
so the agent that writes the code is never the one that judges it.}
\label{fig:green}
\end{figure*}
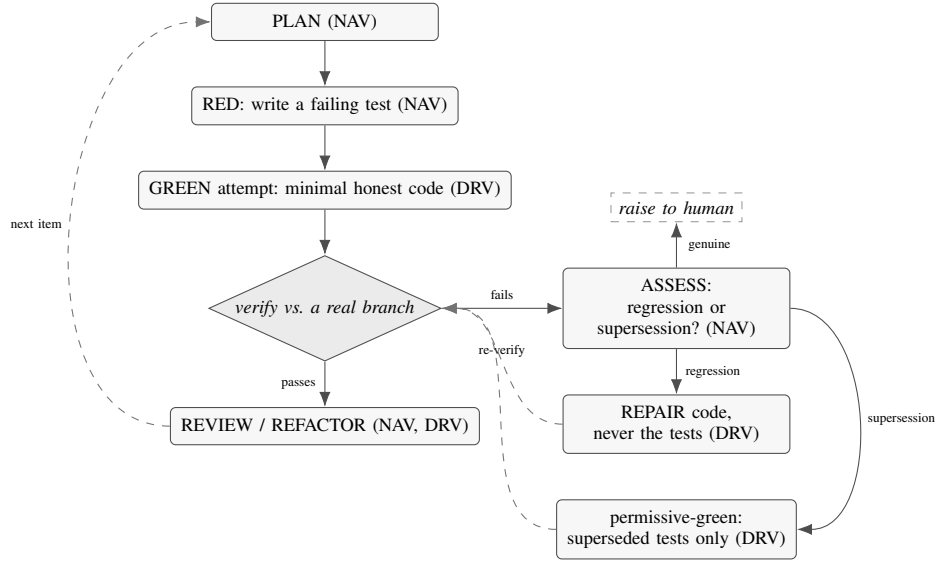

\subsection{The design lane}
Before the enforced build lane runs, the design lane is where Consort meets the
spec-driven field on its own ground. Intent is captured in a specification and a list of
the tests that will demonstrate it, and both are frozen at an approval gate and checked
for drift thereafter, so that the target the agent builds against cannot quietly move.
This is spec-first, like the comparators; the difference is that the frozen specification
and test list feed a build lane whose enforcement is structural, so the discipline does
not evaporate when the agent takes over.

\subsection{Context management}
Agent sessions accumulate context until output quality degrades or the model refuses.
Consort bounds this in the orchestrator. Because roles share no memory and pass
everything through on-disk artifacts, any turn can be reloaded cold, so a fresh start is
always safe. The orchestrator tracks each session's token usage against the model's
window and, when a warm resume would leave less than a configurable fraction free (40
percent by default), starts the next turn fresh. It keeps warm resumption where it is
cheap, reusing the prompt cache across a story's cycles. GSD pursues the same goal with
fresh-context subagents; here it is an enforced property of the deterministic driver, not
a convention. Consort also scopes what each turn sees. Rather than the whole codebase, an
agent is handed a context package with the exact tests it must pass, the relevant
requirements, and where those tests live, which lowers cost and removes the unscoped
wandering that otherwise lets an agent drift.

\section{Comparative evaluation}
\label{sec:eval}
The evaluation is, in the first instance, analytic: the five frameworks are placed
against the dimensions that determine whether an agent's output can be trusted, and the
pattern that emerges is that Consort is the only one whose controls are enforced rather
than requested. Table~\ref{tab:dims} summarizes the comparison; the discussion that
follows evaluates each dimension in turn, draws out the crux, and concedes where the
comparators lead.

\begin{table*}[t]
\centering
\caption{Enforcement dimensions across the five frameworks.}
\label{tab:dims}
\scriptsize
\setlength{\tabcolsep}{3pt}
\renewcommand{\arraystretch}{1.15}
\begin{tabular}{@{}L{1.05cm}L{1.05cm}L{1.5cm}L{1.6cm}L{1.2cm}L{1.35cm}L{1.6cm}L{2.0cm}L{1.55cm}L{2.1cm}@{}}
\toprule
Framework & SDD level & Spec frozen / drift-checked & TDD & Tests immutable & Database in loop & Orchestration & Gates & Role separation & Context management \\
\midrule
\textbf{Consort} & Spec-first & Yes (hashed) & Core, enforced & Yes & Live branched DB & Deterministic (code) & Required (human-approved) & Eight named roles* & Memoryless; resets at 40\% free window \\
GitHub Spec Kit & Spec-first & No (editable) & Optional, ungated & n/a & None & LLM (agent-run) & Advisory (prose) & One agent & Batching + sub-agent delegation \\
obra/\allowbreak{}superpowers & Spec-first & No (model-editable) & Strong rule, unenforced & No & None & LLM control loop & Advisory (soft prose) & Ephemeral subagents & Fresh-context subagents (can lose place) \\
BMAD & Spec-anchored & No (editable) & Optional (post-epic QA) & No & None & Deterministic (bmad-loop) & Code-enforced, post-hoc (coverage/quality) & Five named personas & Self-contained sharded stories \\
GSD & Spec-anchored & No (living) & Absent & n/a & None & LLM (agent loop) & Advisory (convention) & Documented taxonomy (hook-enforced) & Fresh-context subagents + STATE/CONTEXT \\
\bottomrule
\end{tabular}

\vspace{2pt}
{\footnotesize *A role is a bounded function the orchestrator enforces: a role cannot do
another role's job, and roles share no memory. A persona (BMAD) is a named character the
agent role-plays, honored by the model, not an enforced boundary.}
\end{table*}

\subsection{Dimension by dimension}
\textbf{Spec-driven level.} All five frameworks capture intent before the build, and
two, BMAD and GSD, keep it anchored as living artifacts. The rung on the ladder is not
itself an enforcement property: a higher rung buys nothing if the captured intent is not
binding on the agent. What makes intent binding is the next dimension.

\textbf{Spec freeze and drift.} A living specification is one the agent can edit, and
for a non-deterministic author, editable and able-to-launder-a-failure are the same
property: under pressure to pass, the agent can drift the spec to match whatever it
built. Consort freezes the specification and its test list at a hashed gate for the
duration of an increment, so the target cannot move mid-build, and evolves it across
increments under a human re-plan. A living spec is right when a human keeps it honest; a
spec frozen per increment is right when an agent does the editing.

\textbf{Test-driven development.} Tests are the executable form of the spec, so where
they sit decides whether output is checked as it is written. Optional or post-hoc
testing (Spec Kit's ungated tests, BMAD's after-the-epic quality assurance, GSD's absence
of a test concept) leaves the moment of authorship unguarded, and a strong but unenforced
rule (superpowers' Iron Law) depends on the model honoring it. Consort makes the
red/green/refactor cycle the build itself: code that does not satisfy a previously written
test does not come into being.

\textbf{Test immutability.} The test is what judges the code, so if the agent can weaken
or delete it, a green result certifies nothing \cite{beck2025}. Only Consort holds tests
immutable within a unit of work, with a single bounded, re-verified exception for
legitimate supersession (Section~\ref{sec:primitives}). Everywhere else the tests remain
editable by the same agent they are meant to constrain.

\textbf{Database in the loop.} This is the one dimension that changes what the agent is
verified against rather than how its process is run, and it is the subject of
Section~\ref{sec:axis}. No comparator has a database in the loop, so none can surface the
data-dependent defects that only real data exposes.

\textbf{Orchestration.} When the control flow is itself a language model, as it is in all
four comparators, it can be lost across context compaction or reasoned around, so the
process meant to enforce the rules is subject to the same non-determinism as the work it
is meant to control. Consort's routing is deterministic code, so the sequence of steps and
gates cannot itself drift.

\textbf{Gates.} A gate is a control only if it can refuse. Advisory gates, whether prose, a
waivable pass-concerns-fail file, or convention, request compliance and can be passed by
the agent or waived by the team. Consort's gates require a human to approve each decision
before the work advances.

\textbf{Role separation.} A single agent that writes and judges its own work has every
incentive to grade generously, and beyond integrity, focused roles are where
maintainability comes from, reproducing the human division of labor
(Section~\ref{sec:roles}). BMAD's six named personas capture much of this; Consort adds
that a role cannot perform another's job and that the judge runs against real data.

\textbf{Context management.} Every framework must fight context rot, and most do it by
convention: fresh-context subagents invoked by prompt (GSD), self-contained sharded
stories (BMAD). Consort measures it instead, the orchestrator tracks each session against
the model's window and resets to a fresh, disk-reloaded turn once a resume would leave too
little of the window free (Section~\ref{sec:consort}), so the bound is enforced by code
rather than by the agent staying disciplined.

\subsection{The crux}
The table maps onto the three modes of Section~\ref{sec:enforce}. superpowers enforces by
persuasion: its test-first rule, its orchestration, and its tests are all editable by the
model. Spec Kit enforces by front-loaded structure: its specifications are
checklist-validated and its build-time enforcement is optional. GSD sits alongside them,
agent-driven with convention- and hook-level gates and a documented role taxonomy that is
not code-guaranteed.

In the interval between this paper's first draft and its final revision, one comparator
moved further. BMAD's bmad-loop, a deterministic, code-driven orchestrator whose
test-architect gates are code-enforced blocks with human-only waivers, landed in that
window, so deterministic orchestration and code-enforced gating are no longer unique to
Consort. But BMAD's gates enforce coverage and quality thresholds computed after an epic,
not the in-cycle, immutable-test, real-data verification that Consort's gates enforce. On
what is gated, a real test passing against a live, branched database with tests the agent
cannot weaken, Consort remains alone.

Consort enforces through controls the agent cannot edit: the orchestrator is code, a human
gates each decision, the tests are immutable, and green is a real run against a real,
branched database. Where the others request compliance, or enforce it only after the fact,
Consort enforces it at the moment of authorship against real data, and the difference is
not one of degree but of whether a determined or careless agent can proceed without
complying.

\subsection{The uncontested axis}
\label{sec:axis}
One row has no contest at all. None of the comparators runs the test loop against a real,
branched database, because none has a database in the loop: Spec Kit stops at a design-time
data model, and superpowers, BMAD, and GSD have no data concept. We state the novelty
precisely, as that paper does: testing against a database is old, and integration
and acceptance tests have always done it; what is new is running the entire cycle, including
the fast inner loop, against a branch of real data at a per-cycle cost low enough to be
routine \cite{companion}. That axis is where Consort's output claim ultimately rests,
because it is the only one of the enforcement dimensions that changes what the agent is
verified against, not merely how its process is controlled.

\subsection{Where the comparators lead}
The comparators do things Consort does not. Today they are more portable, running across
many coding agents, harnesses, and runtimes. Consort was first developed on Claude, but
will soon be multi-harness and multi-model through the Unity Gateway (planned). Spec Kit
also has an editable constitution that Consort does not support yet. Consort trades
portability and lightness for verifiability of the agent's output: a team that values reach
across many agents, or that is working on code where the maintainability risk is low, may
reasonably prefer a lighter framework. Enforcement through controls the agent cannot edit
carries a cost, warranted when the cost of wrong or unmaintainable agent output is high: it
is the only one of the three modes whose guarantees hold no matter how the agent behaves.

\section{A controlled evaluation (pre-registered, future work)}
\label{sec:experiment}
The analytic comparison establishes that Consort enforces what the others request. It does
not, by itself, establish that enforcement produces better code, and we are careful not to
assert the stronger claim as though it were demonstrated. The output-quality claim is a
hypothesis, and we state it as one.

\begin{tcolorbox}[colback=black!3,colframe=black!55,boxrule=0.5pt,arc=2pt,left=4pt,right=4pt,top=3pt,bottom=3pt]
\emph{Given the same agent and the same feature, a framework with enforced immutable tests,
deterministic gates, real-data verification, and separated role agents yields output with
fewer silent regressions, no test-gaming, and higher maintainability than frameworks that
rely on the model honoring prose rules.}
\end{tcolorbox}

The experiment that would test this holds the agent and the task fixed and varies only the
enforcement regime. The same features, drawn from industrially relevant, stateful tasks
rather than stateless katas, are implemented under each of the frameworks with the same
underlying model. Five families of measure follow the hypothesis. Silent regressions are
counted as defects that pass a framework's own gates but are caught by an independent,
held-out test suite. Test-gaming is counted directly, as tests deleted, weakened, or
adjusted to force a green result, the dimension on which immutability should show its
effect. Maintainability is assessed both by independent review and by objective proxies such
as duplication and specification-to-code drift, the same signals whose decline motivates the
study \cite{gitclear2025}. And real-behavior coverage is measured as the share of tests
exercising a real database rather than doubles, which ties Consort's distinctive axis to an
outcome. Cost is measured as tokens per shipped feature at a fixed model, and is reported
against the quality measures rather than alone, so the comparison is quality gained per token
spent, not lowest cost.

We pre-register the hypotheses and the direction of each effect, and assert no target
magnitudes in advance, so that the analysis answers the rigor-and-relevance concerns that
have dogged tool-comparison studies. Several threats to validity are visible from the outset.
The author builds one of the frameworks under test, a dual role that must be declared and
mitigated with framework-neutral tasks, independent reviewers, and an invitation to replicate.
The frameworks differ in harness portability, so the agent must be held constant across
regimes to keep the comparison about enforcement rather than tooling. And maintainability is a
construct whose operationalization is itself contestable and must be fixed in advance. This is
the same evidentiary posture as that paper's deferred productivity experiment
\cite{companion}: the mechanism is argued now, the outcome is measured later, and the remedy is
offered as a hypothesis rather than a proven cure.

\subsection{The scoring instrument, and keeping it independent}
The measures above (silent-regression rate, test-gaming, dropped intent, real-behavior
coverage) are scored by an LLM-as-judge on a fixed model against a fixed reference, so the bar
does not move with the system under test. Consort ships such an instrument: its evaluation
package, a fixed-model discriminator that scores an output against a recorded reference and
classifies code as equivalent, a superseded shift, a regression, or insufficient, against the
architect's non-functional requirements and design canon. Because that instrument is part of
the system under test, using it unmodified would not be independent. We therefore plan to release the
evaluation package as an open, framework-neutral artifact so the measurement can be inspected
and rerun by others, and would bind it with five safeguards:
\begin{enumerate}
  \item Neutral reference: score every framework against a pre-registered specification and
  acceptance criteria authored independently of any framework, not against one framework's
  recorded artifacts.
  \item Judge decoupled from generator: judge with a model distinct from the one generating the
  code, or an ensemble, and report inter-judge agreement.
  \item Human calibration: validate the judge against expert human ratings on a sample and
  report the agreement.
  \item Blinding: the judge is not told which framework produced an output.
  \item Independent execution: the study is run and scored by parties not authoring the
  frameworks under test.
\end{enumerate}
Released and neutralized this way, the instrument would become a contribution that invites
replication rather than a private grader.

\section{Discussion}
\label{sec:discussion}
Consort's defining choice is to put a live database in the loop and enforce the cycle against
it through controls the agent cannot edit. That requires a database dependency and enforcement
machinery the other frameworks do not, and it is not free. That weight buys guarantees that
hold regardless of how the agent behaves, which matters most where the cost of wrong output is
high: regulated systems, code that touches money or personal data, and long-lived code whose
maintainability compounds. For a throwaway prototype the weight is not worth carrying.

Real-data verification has a particular consequence for regulated work. The reason teams in
sensitive domains test against synthetic data in staging is that production data is masked or
off-limits, and an AI assistant's unverified output does not survive an audit \cite{mammo2025}.
A branch that carries real data with its lineage and access controls intact makes real-data
testing admissible where a raw copy would not, and enforcing the controls in code supplies the
audit trail: the gate that a human approved, the immutable test that judged the code, and the
branch it ran against. The combination is not merely a better inner loop; it is an argument
that agent-written code was verified against the real thing under controls that an auditor can
inspect.

More broadly, the field is converging on the conclusion that a specification without
enforcement is not enough, whether stated as platform enforcement \cite{itential2026}, as
specifications that execute as validation gates \cite{augment2026}, or as the observation that
spec-driven discipline evaporates once the agent begins to write \cite{bockeler2025}. Consort's
position within that convergence is a specific one: put the controls where the agent cannot edit
them, and define success as a real run against real data. As more code is written by agents, the
artifact the agent is held to, an immutable test on a branch, becomes the locus of control,
displacing the prompt. That is the transferable idea, independent of any one framework: enforce
the discipline through controls the agent cannot edit, not by persuasion.

\section{Conclusion}
\label{sec:conclusion}
Spec-first, agent-driven development frameworks have gained rapid traction since 2025, and they
agree on how intent is captured. What separates them is how they enforce clean, correct code
from that intent, and enforcement comes in three strengths: persuasion, which the agent may
ignore; front-loaded structure, which the agent outruns once it starts building; and controls
the agent cannot edit. We have introduced Consort, a spec-first, test-driven agent framework, as
a reference realization of the third mode, whose enforced build lane runs the entire test-driven
cycle against a live, branched database, the practice developed in a paper under anonymous review. We have argued
analytically that it enforces what the others request, and we have stated the output-quality
claim as a pre-registered hypothesis rather than a proven result. The transferable contribution
is the principle, not the product: when an agent writes code, the controls that matter are the
ones it cannot ignore, and the result that matters is the one a real database will actually
accept. We offer Consort and the principle for adoption and independent evaluation.

\end{document}